\documentclass[%
twocolumn,
superscriptaddress,
amsmath,
amssymb,
pra
]{revtex4-1}
\usepackage[dvipsnames]{xcolor}
\usepackage[pagebackref=false]{hyperref}
\usepackage[utf8]{inputenc}
\usepackage{textcomp}
\hypersetup{
    colorlinks  =   true,
    linkcolor   =   BlueViolet,
    citecolor   =   BlueViolet,
    filecolor   =   BlueViolet,
    urlcolor    =   BlueViolet}
\usepackage{graphics,amssymb,amsmath,epsfig,color}
\usepackage{graphicx}
\usepackage{epstopdf}
\usepackage{dcolumn}
\usepackage{bm}
\usepackage{color}
\usepackage{ulem}

\newcommand{\cmt}[1]{}

\begin{document}

\title{A K-band Kinetic Inductance Parametric Amplifier Near the Quantum Limit} 

\author{Chaofan Wang}
\affiliation{Department of Electrical Engineering, Yale University, New Haven, Connecticut 06520, USA}

\author{Shihan Liu}
\affiliation{Department of Electrical Engineering, Yale University, New Haven, Connecticut 06520, USA}

\author{Yufeng Wu}
\affiliation{Department of Electrical Engineering, Yale University, New Haven, Connecticut 06520, USA}

\author{Danqing Wang}
\affiliation{Department of Electrical Engineering, Yale University, New Haven, Connecticut 06520, USA}

\author{Manuel C. C. Pace}
\affiliation{Department of Electrical Engineering, Yale University, New Haven, Connecticut 06520, USA}

\author{Xiangzheng Li}
\affiliation{Department of Electrical Engineering, Yale University, New Haven, Connecticut 06520, USA}

\author{Hong X. Tang}
\email{hong.tang@yale.edu}
\affiliation{Department of Electrical Engineering, Yale University, New Haven, Connecticut 06520, USA}

\date{\today}

\begin{abstract}
Advancing superconducting quantum devices to higher operating frequencies broadens their functionality and enables operation at elevated temperatures, but it also requires near-quantum-limited amplifiers beyond the few-gigahertz regime. Here we present a junction-free, kinetic-inductance parametric amplifier based on thin-film niobium nitride (NbN) operating at 23 GHz in the microwave K-band, achieving a gain up to 40 dB, a 100 MHz gain–bandwidth product, a 1 dB saturation input power of –85 dBm with 23 dB gain, and added noise no greater than 1.4 quanta for phase-preserving amplification. Leveraging the large superconducting gap of NbN, this architecture can be extended to even higher frequencies, supporting applications such as high-fidelity readout of millimeter-wave superconducting qubits and axion searches over an expanded mass window.

\end{abstract}

\maketitle
Superconducting quantum devices, particularly superconducting qubits, have become notable platforms for quantum sensing and quantum information processing \cite{devoret_superconducting_2013,blais_circuit_2021,krantz_quantum_2019}. While most state-of-the-art implementations of these devices operate below 10 GHz, moving to higher frequencies offers significant advantages. High-frequency qubits are inherently less susceptible to thermal excitation, enabling operation at elevated temperatures that are crucial for scalable architecture \cite{anferov_superconducting_2024,anferov_millimeter-wave_2025}. In cosmology, axion dark matter searches are pushing for an increasingly higher frequency range, with a recent interest in the 15-50 GHz frequency range motivated by the SMASH model \cite{quiskamp_direct_2022,quiskamp_exclusion_2024,ballesteros_unifying_2017}.

Progress towards high-frequency platforms, however, is limited by the lack of near-quantum-limited first-stage amplifiers. Without such amplifiers, system performance falls behind that of established lower-frequency technologies. For superconducting qubits, this leads to reduced readout fidelity; in haloscope experiments, it substantially increases the required scan time.

Near-quantum-limited amplifiers are an enabling technology for the high-fidelity readout of weak microwave signals \cite{clerk_introduction_2010,roy_introduction_2016}. Over the past two decades, extensive research has focused on developing such amplifiers for the conventional sub-10 GHz frequency range \cite{castellanos-beltran_amplification_2008,bergeal_phase-preserving_2010,macklin_nearquantum-limited_2015,narla_wireless_2014,white2023readout,joshi_lumped-element_2025,ranadive_travelling-wave_2025}. In contrast, the development of amplifiers for higher frequencies (10 GHz to 300 GHz) remains relatively unexplored.

Two major obstacles limit the development of high-frequency parametric amplifiers. The first is the superconducting gap of commonly used materials. Aluminum, for instance, has a critical temperature $T_c$ slightly above 1 Kelvin, with a superconducting gap $\Delta$ of approximately 40 GHz \cite{tinkham2004introduction}. This intrinsically limits the operating frequency of aluminum-based amplifiers. The second obstacle involves packaging and interconnects. As the frequency increases, parasitic effects from wirebonds and resonant modes in packaging and substrate become more prominent and challenging to mitigate. Consequently, demonstrations of superconducting devices above 20 GHz for quantum information processing have only appeared recently \cite{anferov_millimeter-wave_2020,shu_34ghz_2021,anferov_superconducting_2024,anferov_millimeter-wave_2025,hao2026wireless}.

Kinetic inductance parametric amplifiers (KIPAs) have recently emerged as a promising solution for low-noise amplification \cite{ho_eom_2012_2012,malnou_three-wave_2021,malnou_performance_2022,parker_degenerate_2022,vine_situ_2023,xu_magnetic_2023,faramarzi_48_2024,Reviewer1_KTWPA}. KIPAs fabricated from a single layer material of higher $T_c$ like Niobium Nitride (NbN) or Niobium Titanium Nitride (NbTiN) have demonstrated high gain ($>$20 dB) and near-quantum-limited noise performance at elevated temperature and inside strong magnetic field \cite{malnou_performance_2022,vine_situ_2023,xu_magnetic_2023,vaartjes_strong_2024}. The high critical temperature and design simplicity of NbN KIPAs make them excellent candidates for first-stage amplification in high-frequency quantum systems.

KIPA implementations generally fall into two complementary categories, travelling-wave KIPA and resonance-enhanced KIPA. Travelling-wave KIPAs deliver low noise, high dynamic range amplification over broad bandwidths of a few GHz \cite{malnou_three-wave_2021,malnou_performance_2022,shu_34ghz_2021,Reviewer1_KTWPA}, showing promises for simultaneous readout of multiplexed, frequency-diverse microwave signals \cite{malnou_three-wave_2021,Reviewer1_KTWPA}. Resonance-enhanced KIPAs, on the other hand, leverage resonant field enhancement to achieve generally higher gain, compact device footprint, simplified circuit architecture, and reduced pump-power requirements \cite{parker_degenerate_2022,xu_magnetic_2023}. Their instantaneous linewidth can be sufficiently large for single qubit readout \cite{anferov_superconducting_2024,hao2026wireless} and haloscope experiments \cite{quiskamp_exclusion_2024}. These devices have also enabled electron spin resonance experiments \cite{vine_situ_2023} and squeezed microwave generation \cite{vaartjes_strong_2024}, where simultaneous broad-band frequency coverage is not required. The device demonstrated in this work is a resonance-enhanced KIPA.

In this work, we present a single-layer NbN-based KIPA operating at 23 GHz in the microwave K-band. The device exhibits a parametric gain of over 40 dB, a gain-bandwidth product exceeding 100 MHz, and a high saturation power greater than -85 dBm at 23 dB gain. By calibrating against a next stage high electron mobility transistor (HEMT) amplifier, we establish an upper bound on the added noise of our device at less than 1.4 photons. These combined performance metrics make our KIPA an ideal candidate for integration into K-band qubit readout systems.


\begin{figure}[tb]
    \centering
    \includegraphics[width=\linewidth]{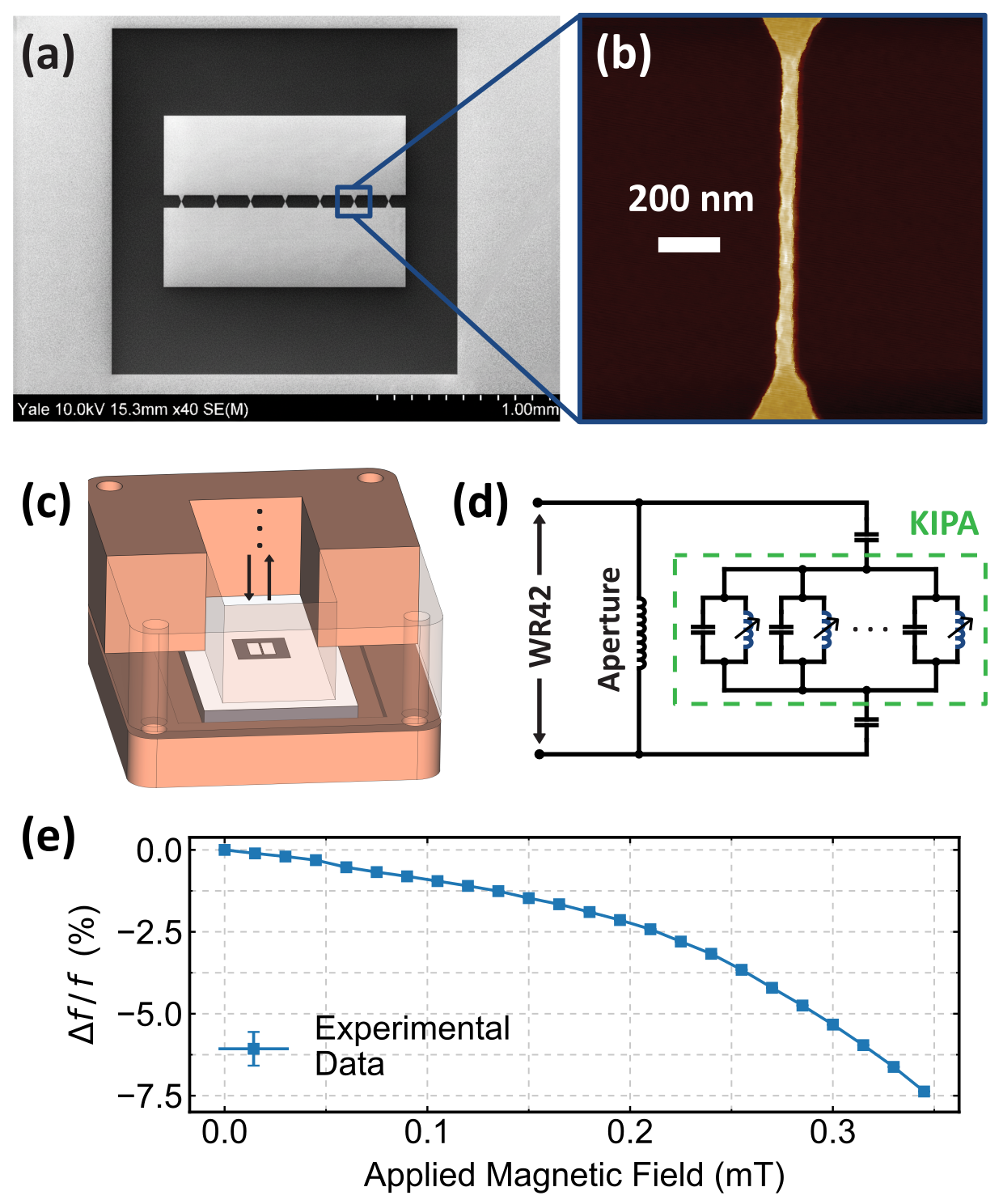}
    \caption{(a) SEM image of the K-band KIPA device, featuring a pair of capacitor pads connected by an array of parallel nanowires. The bright area is NbN, and the dark area is exposed silicon. (b) AFM image of one NbN nanowire. (c) The K-band KIPA device is clamped between a WR-42 waveguide and a recess in the copper base. (d) The equivalent circuit diagram of the KIPA device coupled to the TE10 mode in the WR-42 waveguide. (e) Frequency tuning of the resonator with a magnetic field applied perpendicular to the device plane.}
    \label{fig:schematic}
\end{figure}

The KIPA device comprises of an array of parallel nanowires bridging a pair of capacitor pads. An scanning electron microsopce (SEM) image of the KIPA is shown in Fig.1(\textcolor{black}{a}), and a zoomed-in atomic force microscope (AFM) image on one of the nanowires is shown in Fig.1(\textcolor{black}{b}). The nanowires are \textcolor{black}{50 nm $\times$ 1$\mu$m $\times$ 25 nm} in width, length, and height. The kinetic inductance of a single nanowire is estimated to be \textcolor{black}{300 pH}. Collectively, the set of parallel nanowires contribute about 50\% of the total inductance of the KIPA. 

The KIPA is coupled to a WR-42 waveguide's TE10 mode by clamping the waveguide on the device as shown in Fig.\ref{fig:schematic}(\textcolor{black}{c}). The waveguide mode's E-field direction is parallel to the nanowire direction, aligning with the KIPA's electric dipole. The array of nanowires and their capacitor pads enhance the coupling of the KIPA to the waveguide input field. The equivalent circuit diagram of the KIPA is shown in Fig.\ref{fig:schematic}(\textcolor{black}{d}), where the inductance and capacitance seen by each wire element are placed in parallel. The waveguide coupling scheme situates the device in a simple, single-mode electromagnetic environment, making the device's coupling more predictable and easier to adjust compared with using coaxial probes \cite{narla_wireless_2014,hao2026wireless}.

The KIPA device was fabricated at wafer scale from a single layer of NbN on a high-resistivity 4-inch silicon substrate. The NbN film was grown by atomic layer deposition for 350 cycles, following a recipe similar to that in \cite{cheng_superconducting_2019}. Ellipsometer measurements indicate a film thickness of 25 nm, corresponding to a sheet kinetic inductance of approximately 10 pH/$\square$. The superconducting critical temperature of this film is around 12.5 K, significantly higher than that of aluminum and tantalum, which are the commonly used in Josephson junction devices. The higher $T_c$ enables KIPA operation at elevated temperatures and higher frequencies.

After film deposition, the nanowires are defined using electron-beam lithography with a positive CSAR150 resist, followed by CF$_4$ reactive ion etching (RIE) to pattern the NbN layer. The capacitor pads are subsequently patterned by photolithography and etched using the same CF$_4$ RIE process. The completed devices are screened by room temperature resistance measurement before being diced into individual dies. The KIPA devices' room temperature resistance shows up to $\pm 4$\% span across a 4-inch wafer, with occasional outlier dies that can be filtered out during room-temperature screening. The KIPA devices' resonance frequency at 4K show $\pm 4\%$ span, aligning with the room-temperature resistance measurement. Due to the simplicity of KIPA fabrication, we consistently achieve $>90\%$ yield over a 4-inch wafer.

\begin{figure}[tb]
    \centering
    \includegraphics[width=\linewidth]{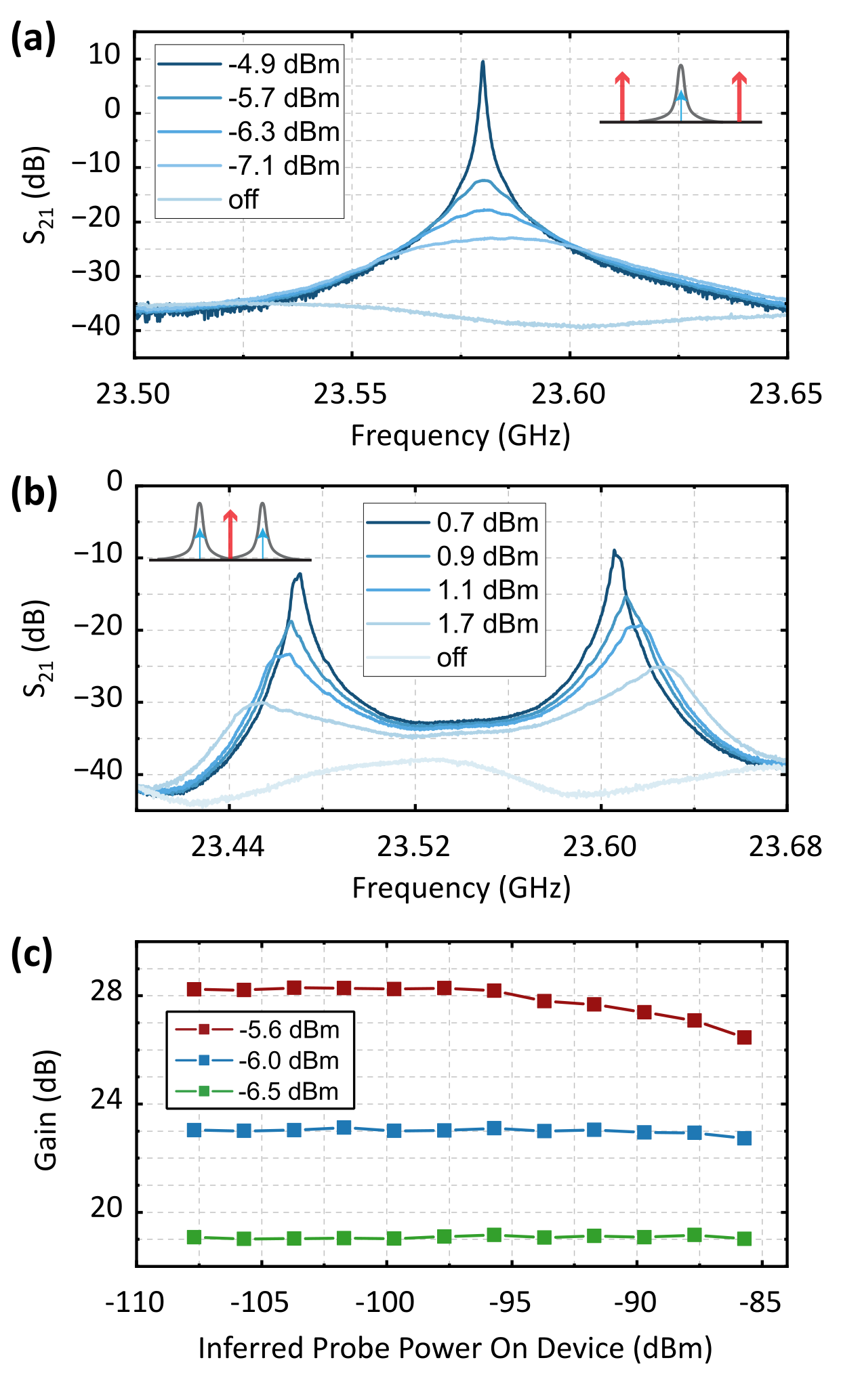}
    \caption{(a) KIPA gain profile under a two-tone pump. The legend corresponds to room-temperature pump power of a single tone. The pump powers are calibrated with a spectrum analyzer (SA) at room temperature (RT), and the total RT-to-mK attenuation of the pump line is conservatively estimated to be 60 dB, with a possible underestimation up to 10~dB. (b) KIPA gain profile under a detuned one-tone pump. (c) Saturation power characterization of the KIPA. The gain value here is referred from on-off measurements at a frequency near the gain peak. }
    \label{fig:gain}
\end{figure}

To test the device, the KIPA is mounted on the mixing chamber plate of a Bluefors dilution refrigerator operating below 20 mK. A directional coupler is used to separate the reflected signal from the input line. In a back-action sensitive setting, a circulator is required in the probe input line, similar to the settings in \cite{hao2026wireless}.

The device exhibits a resonance frequency of \textcolor{black}{23.7 GHz} within the K-band. To tune the resonance frequency, a superconducting coil magnet is placed underneath the device to apply a magnetic field perpendicular to the device plane. The applied field induces a DC current which loops through the first and the last nanowire. The bias current increases the kinetic inductance of these two wires, thereby reducing the resonance frequency of the device. The resonance frequency and gain of the KIPA device remains stable for at least 12 hours of continuous operation under the applied field. The frequency tuning curve is shown in Fig.\ref{fig:schematic}(e). The tuning range of this frequency-tuning scheme is generally limited to a few percent, constrained by the maximum DC current in the nanowires before it becomes prone to vertex tunneling \cite{mingrui_tuning}. Therefore, it is useful for fine frequency tuning within a moderate range, such as aligning with a qubit readout cavity in qubit readout experiments, but not yet for broad-range frequency tuning over multiple GHz. For full-band coverage, the KIPA device can be engineered to support an array of near-uniformly spaced modes in frequency, and use magnetic field tuning to align one of these modes to the desired frequency \cite{yufeng_array}. The external coupling rate of the device is fitted to be \textcolor{black}{$2\pi\times60$ MHz}, which sets an expectation for the gain-bandwidth product of the amplifier.

The dynamics of the KIPA is governed by the Hamiltonian $\hat{H}=\hbar\omega_0\hat{a}^\dagger\hat{a}-\hbar K\hat{a}^\dagger\hat{a}^\dagger\hat{a}\hat{a}$,
where $\hat{a}$ ($\hat{a}^\dagger$) is the annihilation (creation) operator, $\omega_0$ is the resonator's angular 
frequency, $K$ denotes the self-Kerr nonlinearity. For this particular KIPA, we estimate that the Kerr non-linearity is $K\approx 2\pi\times1$ kHz based on the measured bifurcation power. The uncertainty of this estimate is primarily limited by calibration of the on-chip pump power. The pump line attenuation is a lower bound that can underestimate by up to 10 dB, and this uncertainty carries over to $K$ and yields $K=2\pi\times1$ - 10 kHz. The nonlinearity enables four-wave mixing (4WM) gain in the KIPA. One way to introduce the 4WM gain is to pump the KIPA device with two tones, $f_{p1}$ and $f_{p2}$. If both tones are sufficiently detuned from the KIPA resonance $f_r=\omega_0/2\pi$, but their midpoint frequency $f_\text{mid}=(f_{p1}+f_{p2})/2$ is close to $f_r$, the undesired 4WM process of $2f_{p1,2}\rightarrow 2f_\text{signal}$ will be suppressed, while the $f_{p1}+f_{p2}\rightarrow2f_\text{signal}$ process will be resonantly enhanced by the KIPA cavity. Under these conditions, the amplifier exhibits a peak gain at $f_\text{gain}=(f_{p1}+f_{p2})/2$. An illustration of this process is shown in the inset of Fig.~\ref{fig:gain}(a).

In Fig.\ref{fig:gain} (a), we demonstrate parametric gain by the resonantly enhanced $f_{p1}+f_{p2}\rightarrow2f_\text{signal}$ around 23.6 GHz, within the K-band frequency range. Here, the two pumps are detuned from the gain center by $\pm 200$ MHz, which is over two KIPA resonance linewidths. With different pump powers, we can demonstrate 4WM gain up to 40 dB, far exceeding the gain required to saturate the readout signal-to-noise ratio (SNR) when being used as the first stage amplifier in a qubit readout chain. At around 20 dB gain, the 3 dB gain bandwidth is 10 MHz, resulting in a gain-bandwidth product of 100 MHz. This gain-bandwith product allows interfacing with K-band qubit readout cavities \cite{anferov_superconducting_2024,hao2026wireless} and microwave resonators in K-band haloscope experiments \cite{quiskamp_exclusion_2024}.

The gain reported here is extracted from an on-off gain measurement. The on-off gain is not expected to significantly deviate from the true gain as typically observed in travelling-wave-based amplifiers \cite{Reviewer1_KTWPA}. A slight discrepancy remains possible due to potential changes in the impedance matching and device dissipation conditions between the pump-on and pump-off states. A more accurate gain calibration would require a calibrated output line, which we leave for future work.

It is also possible to utilize the inverse process, where a single pump generates two distant gain peaks $2f_{p}\rightarrow f_\text{signal,1}+f_\text{signal,2}$, as shown in Fig.\ref{fig:gain} (b) inset. This inverse process can also be enhanced by the cavity, given either $f_\text{signal,1}$ or $f_\text{signal,2}$ is close to the KIPA resonance frequency $f_r$, and $f_p$ is sufficiently off-resonant so that the two gain peak does not overlap with each other. The benefit of such gain scheme is that it allows us to use a single-mode nonlinear resonator to achieve phase-preserving gain peaks, as opposed to having to design multiple modes \cite{bergeal_phase-preserving_2010}. With this scheme, we can also achieve larger than 20 dB gain as shown in Fig.~\ref{fig:gain} (b).

One important metric of the parametric amplifiers is their saturation power. We benchmark our amplifier's saturation power at different parametric gain, as shown in Fig.\ref{fig:gain} (c). The probe power is inferred assuming 80 dB total attenuation from room temperature to the KIPA input. This includes the 20 dB directional coupler used to combine the probe with the pump. The probe line attenuation may be overestimated by up to 10 dB, resulting in a conservative lower bound on the KIPA saturation power. At 23 dB gain, we observe a 1 dB saturation power of -85 dBm at the device input port. Due to the potential overestimation of the probe-line attenuation, the true saturation power is expected to fall within the range of -85 dBm to -75 dBm.

This saturation power is considerably larger than single-junction Josephson parametric amplifiers such as \cite{bergeal_phase-preserving_2010,narla_wireless_2014,hao2026wireless}, and on par with SQUID-array based parametric amplifiers and traveling wave parametric amplifiers \cite{macklin_nearquantum-limited_2015,white2023readout,ranadive_travelling-wave_2025}. This can be understood from the reduced effective nonlinearity of the device. A single Josephson junction has a relatively large Kerr nonlinearity, which limits the maximum intracavity photon number and restricts the dynamic range. In JJ-based amplifiers, an array of junctions is used to dilute the nonlinearity while preserving the total inductance. Kinetic-inductance nanowires exhibit a weaker and more distributed nonlinearity without strong higher-order terms. In this sense, the effect of a single nanowire is analogous to that of a Josephson junction array, leading to a smaller effective Kerr coefficient and higher saturation power. In addition, our implementation connects multiple nanowires and their capacitor pads in parallel. This increases the total cross-sectional area of the nanowires, which further reduces the device nonlinearity. The larger total capacitor pad area also enhances the capacitive coupling to the waveguide walls, thereby increasing the external coupling rate $\kappa$. The combination lifts the saturation power $P_{\text{sat}}$ according to $P_{\text{sat}}\propto \kappa/|K|$. Having a high saturation power is crucial for high-frequency qubit readout experiments in and beyond K-band. This is because the per-photon energy and readout speed both scale linearly with readout frequency, resulting in a quadratic growth in readout power as the readout frequency scales up.

\begin{figure}[tb]
    \centering
    \includegraphics[width=\linewidth]{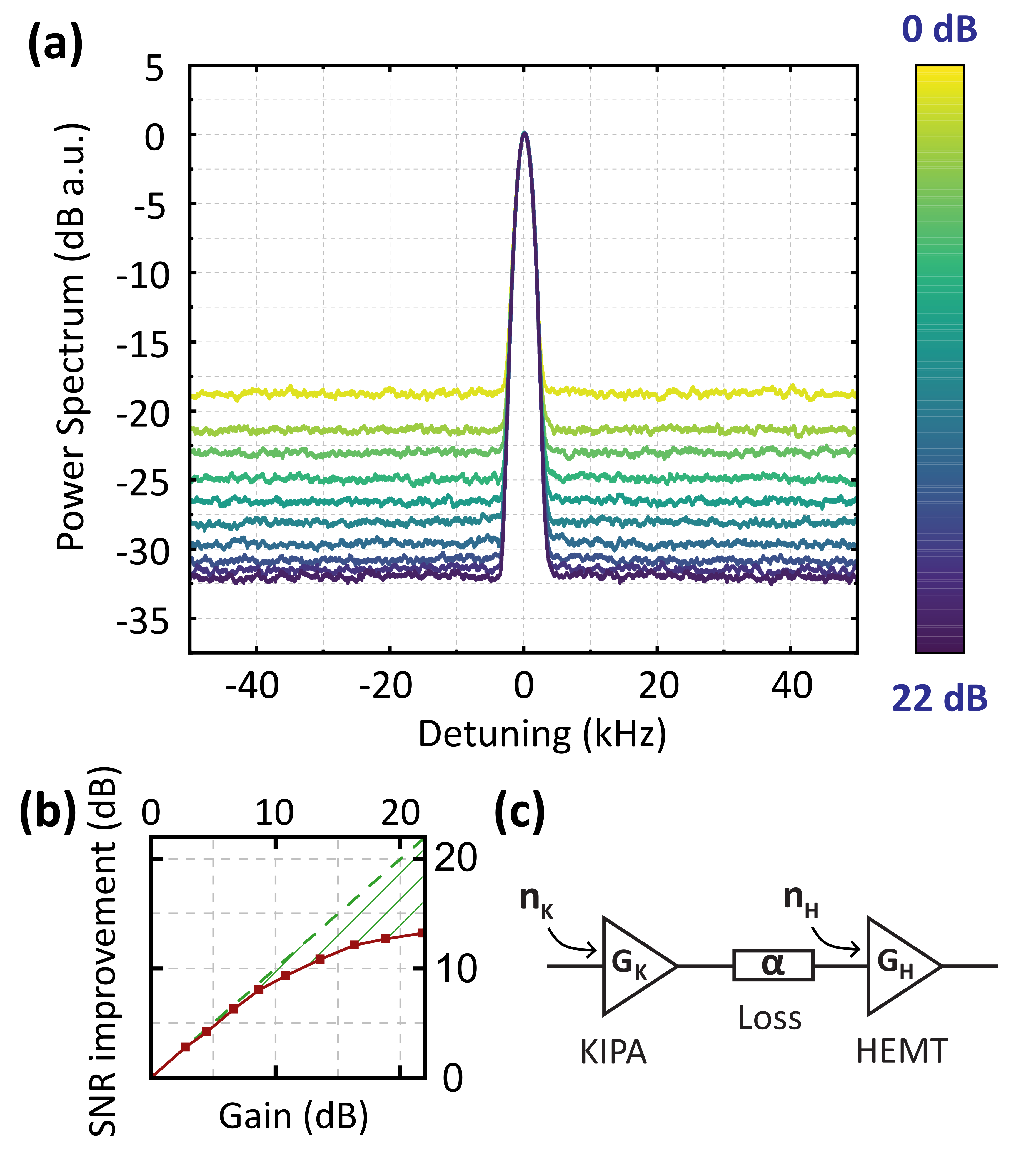}
    \caption{(a) Signal-to-noise ratio measurement at different KIPA gain. Different colors represent different KIPA gains. The power spectrum is normalized to its peak power, such that the improvement of SNR corresponds to the decrease of noise floor. (b) The SNR improvement as a function of KIPA gain. The green dashed line is to guide the eye, showing a linear relationship through the origin. (c) The noise model used for analyzing the SNR improvement.}
    \label{fig:noise}
\end{figure}


To estimate the added noise performance of the device, we calibrate the amplifier's noise temperature against the next stage amplifier, which is the HEMT from Low Noise Factory. We do this by monitoring the SNR change of the output signal as a function of the KIPA gain, as shown in Fig.\ref{fig:noise} (a) and (b). The model used to infer the KIPA added noise is shown in Fig.\ref{fig:noise} \textcolor{black}{(c)}. With this model, we can give an upper bound of the KIPA added noise, $n_K$, by assuming an upper-bound of the loss $\alpha$ between the KIPA output and the HEMT input.

At the input of the KIPA, we send a weak single-tone probe signal significantly below the device saturation input power. The input line is heavily attenuated from RT to MXC, such that the noise of the input tone is sufficiently thermalized with the MXC environmental noise.

We then change the KIPA's pump condition to enable a variable gain from 0 to 22 dB, and measure the ouput power at room temperature with a spectrum analyzer. The result is shown in Fig.~\ref{fig:noise}(a). The data was taken at a resolution bandwidth of \textcolor{black}{2 kHz} and video bandwidth of \textcolor{black}{1 Hz}. By changing the RBW and probe signal power, we make sure the noise floor we see in Fig.~\ref{fig:noise}(a) is not due to the phase noise of the probe tone. 

In Fig.~\ref{fig:noise}(a), we normalize the data with respect to the peak amplified power. In this case, the peak powers overlap at 0 dB a.u., and the increased SNR is reflected in the lowering of the noise floor. In Fig.\ref{fig:noise} (b) we plot the SNR improvement with respect to the SNR when the KIPA is not pumped. The SNR improvement saturates at high KIPA gain, which is a result from the KIPA-amplified noise saturating the HEMT added noise. This yields information on the KIPA added noise.

Quantitatively, we model the SNR improvement factor as:

\begin{align}
        \text{SNR}_\text{improve}=\frac{n_0 +n_\text{post}}{n_0+n_K+n_\text{post}/G_K}
    \label{eq:SNR_improve}
\end{align}
where $n_0=0.5$ is the noise floor of the probe signal, assuming the input line is well-thermalized with the MXC environment;  $n_K, G_K$ are the added noise and the gain of the KIPA respectively; and $n_\text{post}=\left((1-\alpha)n_T + n_H\right)/\alpha$ is the total noise added after KIPA (by the loss bath and HEMT), referred to the KIPA input.  Here, $\alpha$ is the attenuation between the KIPA output and HEMT input, which is assumed to be at mili-kelvin with environmental noise photon $n_T=0.5$; $n_H$ is the added noise of the HEMT. 

Equation \eqref{eq:SNR_improve} assumes the amplified noise of the HEMT saturates all the other noise sources after the HEMT. This is justified by the high HEMT gain of \textcolor{black}{26 dB}. We look for the point $G_K=G_{K,-3\text{ dB}}$, where $\text{SNR}_\text{improve}/G_K=-3\text{ dB}$. This provides information on the KIPA's added noise.

From the SNR improvement measurement, we can extract $G_{K,-3\text{ dB}}=\textcolor{black}{\text{14 dB}}$. Additionally, we can give an upperbound for the KIPA-to-HEMT loss $\alpha\leq -7\text{ dB}$. See Table S1 in the Supplementary Material for an accounting of the losses. We also estimate the upperbound of the HEMT noise to be $n_H\leq 9$, which corresponds to HEMT noise temperature $T_H\leq 10\text{ K}$. 

Using these numbers, we can derive an upper-bound for our KIPA added noise:

\begin{align}
    n_K=\frac{2n_0+ n_\text{post}}{G_{K,-3\text{ dB}}}-n_0\leq n_{K,\text{max}}=1.4
    \label{eq:added_noise}
\end{align}
which is about \textcolor{black}{3} times the quantum limit. This estimate is obtained using a simplified noise calibration approach, which yields an upper bound on the added noise. The sensitivity of the extracted noise to the estimated interstage loss is shown in Fig. S1 of the Supplementary Material. A calibrated noise source, such as a variable temperature stage (VTS) \cite{xu_magnetic_2023} or a shot-noise tunneling junction (SNTJ) \cite{JPAarray_SNT,Reviewer1_KTWPA}, is required to refine this measurement in future experiments. However, these technologies are not yet immediately available for K-band measurements: waveguide-coupled VTSs are difficult to thermally isolate from our circuit, while SNTJs are not yet available at K-band frequency. Another possible noise calibration technique involves using a transmon-cavity system whose readout cavity is in the K-band \cite{hao2026wireless}. However, readout with large detuning or alternatively a higher frequency qubit are still areas of active research. Pushing one or more of these noise calibration techniques to K-band and beyond will be a critical step for our future work.

Our future work would also include operating our K-band KIPA at elevated temperatures of 1 K and beyond. Following a previous demonstration of radiatively-cooled NbN KIPA at 1 K \cite{yufeng_radiative_cooled}, we expect our K-band KIPA to maintain its added noise, gain, and saturation power performance at 1 K. Our NbN resonator's quality factors remain roughly constant from 20 mK to 1 K, due to our NbN film's superconducting temperature around $12.5$ K. At $T=1$ K, the environment's 23 GHz photon mode will be populated by $n_T\approx1.0$. Assuming the amplifier input line remains thermalized to millikelvin temperatures and the device is over-coupled by a factor of 10, the photon population of the idler mode around 23 GHz will be $n_I\approx0.55$, close to the vacuum level of $n_0=0.5$. Therefore, operating at 1 K is not expected to substantially degraded the KIPA's added noise performance \cite{yufeng_radiative_cooled}.

To conclude, we have demonstrated a NbN kinetic-inductance parametric amplifier operating in the K-band, providing up to 40 dB gain, a gain–bandwidth product of 100 MHz, high 1 dB saturation input power up to -85 dBm, and added noise close to the quantum limit ($\leq$1.4 photons). In a cryogenic readout chain, the device provides a clear improvement in signal-to-noise ratio (SNR), demonstrating its potential for high-frequency superconducting-qubit readout and dark-matter or rare-event searches. The wafer-scale fabrication process and simple single-layer material stack further support scalability and manufacturability. With continued optimization of the packaging and connections between the KIPA and the HEMT amplifier, we anticipate that the KIPA can approach quantum-limited noise performance over broader instantaneous bandwidths.The architecture can naturally extend toward V-band, W-band, and beyond, enabling scalable high-frequency microwave quantum information platforms.

\section*{Supplementary Material}

See the supplementary material for additional details on the estimated interstage loss between the KIPA and HEMT amplifier and the dependence of the inferred KIPA-added noise on the assumed loss.

\medskip
Acknowledgments. This research was sponsored by DARPA under grant no. HR0011-24-2-0346. HXT acknowledges funding by the Army Research Office under Grant No. W911NF-24-2-0240, the Air Force Office of Sponsored Research under Grant No. FA9550-23-1-0688 and  the Office of Naval Research (ONR) under award number N00014-23-1-2121, and the U.S. Department of Energy (DoE), Office of Science, National Quantum Information Science Research Centers,
Co-design Center for Quantum Advantage (C2QA) under contract number DE-SC0012704. The authors would like to thank Yong Sun, Lauren McCabe, Kelly Woods, Yeongjae Shin, Michael Rooks, and Sungwoo Sohn for their assistance in device fabrication. Fabrication facilities use was supported by the Yale Institute for Nanoscience and Quantum Engineering (YINQE) and the Yale Univeristy Cleanroom.

\bibliography{reference}

\clearpage
\onecolumngrid

\setcounter{figure}{0}
\setcounter{table}{0}
\setcounter{equation}{0}
\renewcommand{\thefigure}{S\arabic{figure}}
\renewcommand{\thetable}{S\arabic{table}}
\renewcommand{\theequation}{S\arabic{equation}}

\section*{Supplemental Material for ``A K-band Kinetic Inductance Parametric Amplifier Near the Quantum Limit''}
\addcontentsline{toc}{section}{Supplemental Material}

\section{Estimated Loss between KIPA Device and HEMT}

\begin{table*}[ht]
\centering
\footnotesize
\begin{tabular}{|c|c|c|c|c|c|c|}
\hline
Component & Device Package & Directional Coupler & Isolator & Superconducting Cable & Connectors & \textbf{Total}\\ \hline
Number & 1 & 1 & 1 & 1 & 2 &\\\hline
 Nominal Loss (dB)& 1& 1.1& 1.5& 0.5&1&\textbf{5.1}\\\hline
Upper Bound Loss (dB)& 1.5& 1.5& 2 & 0.5 & 1.5&\textbf{7}\\ \hline
\end{tabular}
\caption{Loss budget for noise estimation.}
\label{tab:loss_budget}
\end{table*}

The device package was developed in-house and its insertion loss was measured using a VNA. The losses of the remaining commercial 2.92-mm connectorized components were estimated from manufacturer specifications. From this analysis, we estimate a nominal interstage loss of approximately 5.1~dB and a conservative upper bound of 7~dB.
\section{Dependence of inferred KIPA noise on loss estimate}
\begin{figure}[ht]
    \centering
    \includegraphics[width=0.7\linewidth]{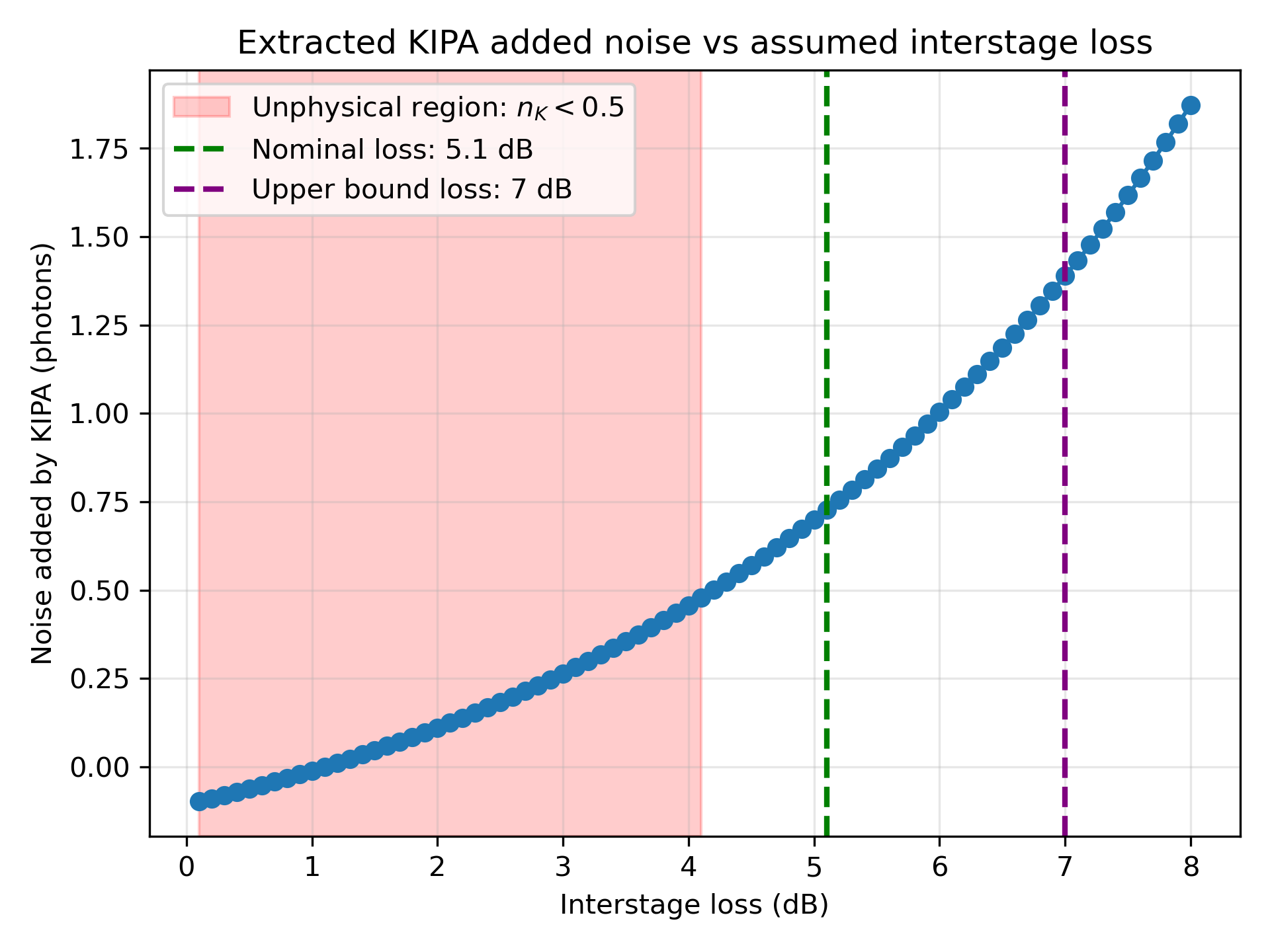}
    \caption{Sensitivity of the extracted KIPA-added noise to the assumed interstage loss.}
    \label{fig:S_loss_sensitivity}
\end{figure}

The extracted noise added by KIPA monotonically increases with more assumed interstage loss.  Therefore, a loss upper bound places an upper bound for the added noise. For the nominal interstage loss of -5.1 dB, the analysis yields 0.7 quanta of noise added by the KIPA. The figure uses the observed $G_{K, -3dB}=14$ dB, which makes it specific to the device and the experimental setup, and not a universal plot. For losses less than -4 dB, the noise added is less than half a photon, which is unphysical in the phase preserving regime. This implies that the interstage loss has to be larger than -4 dB. Otherwise, we would have observed a smaller $G_{K, -3dB}$ (less than observed 14 dB) to begin saturating the SNR improvement.

\end{document}